\documentstyle[12pt]{article}
\newcommand{\beq}{\begin{equation}}
\newcommand{\eeq}{\end{equation}}
\newcommand{\ds}{\displaystyle}

\begin{document}    \begin{flushright}  PITHA 97/18
  \end{flushright}  \begin{center} {\LARGE Canonical  Quantum
 Statistics
  \\[0.1cm] of an Isolated Schwarzschild Black Hole \\[0.1cm]
  with a Spectrum  $E_n = \sigma \sqrt{n}E_P$  \\[0,5cm]
}
{\large H.A. Kastrup\footnote{E-Mail: kastrup@physik.rwth-aachen.de} \\
Institute for Theoretical Physics, RWTH Aachen \\[0.2cm] 52056 Aachen,
 Germany}
\end{center} \vspace*{1.0cm}
  {\large \bf Abstract} \\[0.2cm]
  Many authors - beginning with Bekenstein - have suggested
  that the energy levels
  $E_n$ of a quantized isolated Schwarzschild black hole have the form
  $E_n=\sigma \sqrt{n}E_P, n=1,2, \ldots, \sigma = O(1), $
   with degeneracies
  $g^{n}$. \\ In the present paper  properties of  a system with such a
  spectrum, considered
   as a quantum canonical
  ensemble, are discussed: \\ Its
   canonical partition function $Z(g, \beta =1/k_B T)$,
   defined as a series
  for $g < 1$, obeys the 1-dimensional heat
  equation. It  may be extended  to values $g>1$ by means
  of an integral representation which reveals a cut of $Z(g,\beta)$ in the
  complex g-plane from $g=1$ to $ g \rightarrow \infty$.
   Approaching the cut from
  above yields a real and an imaginary part of $Z$. Very surprisingly, it is
  the (explicitly known) imaginary part which gives the expected
   thermodynamical properties of
  Schwarzschild black holes: \\
   Identifying the internal
  energy $U$ with the rest energy $M c^2$ requires
  $\beta$ to have the value (in natural units) \[ \beta=2M (\ln g /\sigma^2)
  [1+ O(1/M^2)] \] ($4\pi\sigma^2 =\ln g$ gives Hawking's $\beta_H$)
    and yields the
  entropy \[ S=[\ln g/(4\pi \sigma^2)]\, A/4 +O(\ln A)\, ,\]
  where $A$ is the area of the horizon.

  \newpage
   \section{Introduction} As the (Hawking) temperature \beq
   k_B T_H= \frac{E_P^2}{8 \pi Mc^2}~\eeq ($E_P= \sqrt{c^5\hbar/G}$ is
   Planck's energy) of the radiation
   emitted by a black hole\cite{1} is proportional to Planck's constant
   $\hbar$, i.e.\ a quantum effect, it was clear from the beginning that its
   deeper understanding would require a quantum theory of the  gravitational
   properties of black holes. Despite the possible lack of a
    convincing general quantum
   theory of gravity many attempts have been made to identify the quantum
   energy levels $E_n$ of an isolated Schwarzschild black hole.
    Bekenstein\cite{2}
   was the first  to use Bohr-Sommerfeld type quantisation arguments and
   suggested a spectrum  \beq E_n = \sigma \sqrt{n}E_P\;, n=1,2,\ldots, \eeq
   where $\sigma$ is a (model dependent) dimensionless constant of order 1.
  \\ Since then a number of
   authors[3-23]
 have given different arguments for a quantum black hole
   spectrum of the type (2). \\ I myself discussed in ref.\ [19] how such a
   spectrum may be understood in the framework of a stringent canonical
   quantisation of the purely gravitational Schwarzschild spherically
   symmetric system\cite{24}\cite{25}. \\ In the following I shall take the
   relation (2) for granted and ask what its implications for the
   thermodynamics of the system are if  this is viewed as a canonical
    ensemble.
   Again I shall not enter into a discussion of possible conceptual problems
   of such a statistical framework in the context of black
   holes[26-29,31] and shall
   make a few comments after the results have been presented. \\ As to the
   present status of the thermodynamics of black holes see the reviews by
   Wald\cite{29,30}, Brout et al.\cite{31} and Sorkin\cite{32}.
   The present situation as to the quantum states of black
    holes in the framework of string theories has recently been reviewed by
    Horowitz\cite{33}. (Strings can have a spectrum $m_n^2 \propto n$, too!)
   \section{Canonical partition function}
   Following the recent discussion by Bekenstein and Mukhanov\cite{15} (and
   that by the same and other authors before) I assume
   the degeneries $d_n$ of the levels (2) to be $ g^{n}$ (those authors
    actually take
  $d_n =2^{n-1}$ which is, however, not essential for the arguments
   below). It
  is convenient in the following to define $t=\ln g$. \\
   The canonical partition function of the system is   \beq Z(t,x) =
   \sum_{n=0}^{\infty}e^{\ds n t}e^{\ds -\sqrt{n}x}=1+e^{\ds t}
   \tilde{Z}(t,x)~,~~t= \ln g~,~
    x= \beta \sigma E_P~~,\eeq
  where $\tilde{Z}$ is the partition function corresponding to the
   assumptions of Bekenstein and
   Mu\-khanov. \\ The series (3) is obviously
   convergent for $t<0 ~ (|g|<1)$ and converges for $ t=0~ (g=1), x > 0,$
    according to the Maclaurin-Cauchy integral
   criterium\cite{34}. For $t > 0$ the series is divergent, but the function
   $Z(t,x)$ can nevertheless be defined by  continuation (see
   below). \\ The series (3) obeys the heat equation \beq \partial_t Z =
   \partial^2_x Z ~, \eeq  with the boundary values \begin{eqnarray}
   Z(t\rightarrow -\infty, x)&=& 1~~,~~ Z(t=0, x)= \sum_{n=0}^{\infty}
    e^{\ds -\sqrt{n}x}
   \equiv \phi(x)~, \\ Z(t, x \rightarrow \infty) &=& 1~~, ~~ Z(t\neq 0, x=0)
   = \frac{1}{1- e^{\ds t}} \equiv \eta(t) ~~. \end{eqnarray} For the
    series $\phi(x)$ we have the lower
   and upper bounds \beq \int_1^{\infty}d\nu e^{\ds
   -\sqrt{\nu}x}=2(\frac{1}{x}+\frac{1}{x^2})e^{\ds -x}\le \phi(x) -1 \le
   \int_0^{\infty}d\nu  e^{\ds
   -\sqrt{\nu}x}= \frac{2}{x^2}~~. \eeq Observing the behaviour of
   $e^{\ds -\sqrt{n}x}$ between $n= k^2$  and $ n=(k+1)^2,(k+1)^2-k^2 = 2k+1$
    one can sharpen these bounds\cite{35}: \begin{eqnarray}
   (\sum_{k=1}^{\infty}2ke^{\ds -kx}) -e^{\ds -x}& =&(\frac{2}{(1-e^{\ds
   -x})^2}-1)e^{\ds -x}\\ & \le &
   \phi(x)-1 \le \sum_{k=1}^{\infty} 2k e^{\ds -kx}= \frac{2e^{\ds
    -x}}{(1-e^{\ds
   -x})^2}~~. \nonumber \end{eqnarray} As to the physics of the
    system  one is interested in
   properties of the partition function for $t>0$. At a first sight one might
   consider to solve the heat equation for $t\ge 0, x\ge 0$ with the
    known initial
   values (5) and (6) in a standard manner\cite{36,37}: \beq Z(t,x)=
   \int_0^{\infty}dy [K(t,x-y)-K(t,x+y)]\phi(y)
   +\int_0^t d\tau \hat{K}(t-\tau,x)\eta(\tau)~,\eeq where $K$ is the "heat
   kernel" \beq K(t,x)= \frac{1}{\sqrt{4\pi t}} e^{\ds -x^2/(4t)} \eeq and
   $\hat{K}$, essentially, its $x$-derivative, \beq \hat{K}(t,x) =
   \frac{x}{2\sqrt{\pi t^3}}e^{\ds -x^2/(4t)}=-2\partial_x K(t,x)~.\eeq This
   approach does not appear to work, however, because the functions $\phi$
    and $\eta$ are not
   "decent" enough\cite{36,37}: $\eta$ becomes singular as $1/t$ for
   $t\rightarrow 0$ and $\phi$ behaves like $2/x^2$ for $x\rightarrow 0$. The
   latter property can  be inferred from the inequalities (7) and (8).
    \\ An extension of the
   function $Z(t,x)$ - actually it was $\tilde{Z}$ -
   defined by the  series (3) into the complex $g-$plane  was discussed
    100 years ago by the mathematician Lerch\cite{38}:
  Using the relation\cite{39} \begin{eqnarray}
  e^{\ds -\sqrt{nx^2}}& =& \frac{|x|}{\sqrt{\pi}}\int_0^{\infty}dv
  e^{\ds -x^2v^2/4-n/v^2}\\ &=&
  \frac{|x|}{2\sqrt{\pi}}\int_0^{\infty}\frac{\ds d\tau}{\ds \tau^{3/2}}
  e^{\ds -x^2/(4\tau)-n\tau}= \int_0^{\infty}d\tau\hat{K}(\tau, x)
  e^{\ds -n\tau} \nonumber \end{eqnarray} converts the series (3)
   into a geometrical
  one which can be summed under the integral for $t<0, x>0$:
    \begin{eqnarray} Z(t=\ln g,x)&=& \frac{x}{\sqrt{\pi}}\int_0^{\infty}dv
  e^{\ds -x^2v^2/4}\frac{1}{\ds 1-e^{\ds (t-1/v^2)}}\\
  &=& \frac{x}{2\sqrt{\pi}}\int_0^{\infty}\frac{d\tau}{\tau^{3/2}}
  e^{\ds -x^2/(4\tau)}\frac{1}{1-e^{\ds (t-\tau)}}\\ & =&
  \frac{x}{2\sqrt{\pi}}\int_1^{\infty}du
  \frac{\ds e^{\ds -x^2/(4\ln u)}}{\ln^{3/2}u}\frac{1}{u-g}~~. \end{eqnarray}
  Notice that the relation (14) may also be written as \beq
  Z(t,x)=\int_0^{\infty}\hat{K}(\tau,x)\eta(t-\tau)=
\int_{-\infty}^td\tilde{\tau}\hat{K}(t-\tilde{\tau},x)\eta(\tilde{\tau})~,
\eeq
  where $\tilde{\tau}=t-\tau$. \\ Observing that \[ 1=
    \frac{x}{\sqrt{\pi}}\int_0^{\infty}dv
  e^{\ds -x^2v^2/4} \] we get \begin{eqnarray} Z-1=e^t\tilde{Z}(t,x)&=& e^t
   \frac{x}{\sqrt{\pi}}\int_0^{\infty}dv
  e^{\ds -x^2v^2/4}\frac{e^{\ds -1/v^2}}{\ds 1-e^{\ds (t-1/v^2)}}\\ & =& e^t
  \int_0^{\infty}d\tau \hat{K}(\tau,x)e^{\ds -\tau}\eta(t-\tau)~.
\nonumber \end{eqnarray}
 \\ The expressions (13) etc.\  may also be obtained by
  inserting the  representation (12) of $\exp(-\sqrt{n}x)$ into
  the series (3) and Borel summing\cite{40} it for $t<0$. \\
   The integrals converge for all values of
  $g\neq 1, t\neq 0$ (the ones representing $\tilde{Z}$ converge even better
  than those for $Z$). For $real$ $g>1$$(t>0)$ one has to take the
   principal value of
  the integrals. \\
   The integral representations
  (13)-(15) for $Z(t,x)$ are
  solutions
  of the heat equation for all (even complex) $t\neq 0$ as can be
   seen immediately,
  e.g., by replacing the differentiation of $1/(1-e^{\ds t-\tau})$
   in eq.\ (14) with respect to $t$
  by the negative one with respect to $\tau$ and performing a partial
  integration afterwards, or from the eq.\ (16) directly, because
  $\hat{K}(t,x)$ is a solution of the heat equation. \\
  Notice that $\tilde{Z}(t,x)$  is not a solution of the heat equation (4),
  only $e^t\tilde{Z}(t,x)$ is one. \\
   As \beq   Z(t, \lambda x)= \frac{x}{\sqrt{\pi}}\int_0^{\infty}dv
  e^{\ds -x^2v^2/4}\frac{1}{\ds 1-e^{\ds t} e^{\ds
   -\lambda^2/v^2}}~~,~~\lambda >0~,  \eeq  we see
  that \beq \lim_{\ds \lambda \rightarrow 0} Z(t,
   \lambda x)=\frac{1}{1-e^{\ds
  t}}~~,
  \lim_{\ds \lambda \rightarrow \infty} Z(t, \lambda x) = 1 ~~, \eeq in
  accordance with eqs.\ (6), but now for $t>0$. \\
  Correspondingly we get for $\tilde{Z}(t, x)$: \beq
  \lim_{\ds \lambda \rightarrow 0} \tilde{Z}(t, \lambda x)=\frac{1}{1-e^{\ds
  t}}~~,
  \lim_{\ds \lambda \rightarrow \infty}\tilde{Z}(t, \lambda x)
   = e^{\ds -\lambda x} ~~.
  \eeq
 The expressions (13)-(15) or (17) can be used to extend the
  function $Z(t=\ln g,x)$
 or $\tilde{Z}(t,x)$ into the complex $g$- or $t$-planes:
 \\ According to eq.\ (15) $Z(g,x)$ has a branch cut
   in the complex $g$-plane  along
  the real axis from $1$ to $\infty$. The discontinuity of $Z$ across
   the  cut  is given
  by\cite{38,41}
  \beq \lim_{\ds \epsilon\rightarrow
  0^+}[Z(g+i\epsilon)-Z(g-i\epsilon)]= 2\pi i \hat{K}(t,x)~,~~
   \eeq and
  $Z(g,x)$ is an analytic function of $g$ except for this
  cut\cite{38,41}.  \\
  If one, therefore, approaches the cut along the real axis from above,
   the limit \[
 \lim_{\ds \epsilon\rightarrow
  0^+}Z(g+i\epsilon,x),~ g>1~,\] is no longer a real-valued function of $g$
   but has a nonvanishing
  imaginary part $Z_i(t,x)$. Standard procedures used in the
   field of dispersion
  relations\cite{41} yield
  as the real part $Z_r(t,x)$ the principal value integral \beq
  Z_r(t,x)= \mbox{p.v.} \int_0^{\infty}d\tau\, \hat{K}(\tau,x)
  \eta(t-\tau)~, \eeq
   and for the
  imaginary part $Z_i$ the expression \beq Z_i(t,x) = \pi
  \hat{K}(t,x)=\frac{\sqrt{\pi}x}{2t^{\ds 3/2}} e^{\ds -x^2/(4t)}~~. \eeq
  Obviously $Z_r$ and $Z_i$ are solutions of the heat equation seperately.
  $Z_r(g,\cdot)$ is the Hilbert transform\cite{41,42} of $\hat{K}(g,
   \cdot)$.\\
  We shall see that, strangely enough, it is this imaginary part of the
  partition function which gives exactly the thermodynamical
   properties expected for
  black holes! \\ The formal reasons for this can be seen from the
   limits (19)
  which obviously are those of $Z_r$: The limit of $Z_r(t,x)$
   for $x \rightarrow 0$ is {\it negative} for $t>0$, whereas $Z_r$ is
   positive for large $x$! Notice that the integrand in eq.\ (22) is
   negative for $t>\tau  >0$ and that, therefore, $Z_r$ may be negative for
   small $x$. A negative partition function can, however, hardly be
   interpreted thermodynamically where the logarithm has to be taken.
   On the other hand the imaginary part (23) is positive for all $t>0, x>0$
   and does not have the "sign-desease" of $Z_r$!
  \section{Thermodynamics} In the thermodynamics of black holes one is
  especially interested in the behaviour of the system for large
   $\beta$ (low
  temperatures), because the inverse Hawking temperature $\beta_H=1/(k_B
   T_H)$
   is very large for macroscopic black holes (see eq.\ (1)).
  \\ According to eq.\ (20) the real part $\tilde{Z}_r$ of
   the partition function $\tilde{Z}(t,x)$ behaves
  like $ \exp(-x)$ for large $x$ and therefore the
  associated internal energy $U= - \partial \ln Z_r/\partial \beta$ is just
  $E_P$, i.e.\ the lowest possible energy level, as one would expect naively
  from the conventional paradigms of statistical mechanics. \\The main reason
  for not using the real part $Z_r$, however, is its property to become
  negative for small $x$ if $t>0$ as discussed above in connection with eq.\
  (22).\\
  The situation becomes surprisingly interesting and unconventional
  if we take the
  imaginary part $Z_i$, eq.\ (23), as the partition function
  for calculating the thermodynamical
  properties of the system:
  \\ We first calculate the internal energy \[
  U =
  \bar{E}=
  -\frac{\partial \ln Z_i}{\partial \beta}=-\frac{\partial
  Z_i /\partial \beta}{Z_i}=Mc^2~~, \] which we identify
  with the total rest energy $Mc^2$ of the black hole and get
     \beq
  U= \frac{\sigma^2 E_P^2}{2 t} \beta -\frac{1}{\beta} =Mc^2 ~~. \eeq
Solving this equation for $\beta$ and discarding the negative root gives
 \beq \beta =
\frac{Mc^2t}{\sigma^2E_P^2}[1+(1+\frac{2\sigma^2E_P^2}{M^2c^4t})^{1/2}]~~,
\eeq which for $Mc^2 \gg E_P$ leads to
\beq \beta= \frac{2t Mc^2}{\sigma^2 E_P^2}(1+\frac{\sigma^2E_P^2}{2t
 M^2c^4})~.~
 \eeq
For $\sigma^2 =t/(4\pi)$ we get  Hawking's $\beta_H$,
 plus a small correction of order $(E_P^2/(M^2c^4))\beta_H$! \\
Furthermore, for the average of  the energies squared we have \beq
\overline{E^2}=\frac{\partial^2
Z_i}{\partial x^2}\sigma^2 E_P^2/Z_i = \partial_t Z_i\sigma^2E_P^2/Z_i =
  (\frac{x^2}{4t^2}-\frac{3}{2t})\sigma^2E_P^2 ~,\eeq  yielding
  the mean square fluctuations \beq \overline{E^2}-\bar{E}^2=
  -\frac{\sigma^2E_P^2}{2t}
  -\frac{1}{\beta^2}~. \eeq  We see that for $t>0$ the right hand side is
  negative, as expected, corresponding to a negative specific heat.
Considering the fact that the left hand side of eq.\ (28) is the
 difference of two
very large numbers, both of the order $M^2c^4$, the mean  square
 fluctuations are actually very small and essentially universal,
  because the right
  hand side depends on $M$ itself only through the term
  $1/\beta^2$, which is negligible for very large $\beta$! \\
 If we define the average level  number $\bar{N}$ by (see eq.\ (3)) \beq
  \bar{N}=
   \frac{\partial^2 Z_i}{\partial x^2}/Z_i~,\eeq then  we get
 \beq \bar{N}= (\frac{x^2}{4t^2}-\frac{3}{2t})
 = \overline{{E^2}}/(\sigma^2E_P^2)\approx (Mc^2)^2/(\sigma^2E_P^2)~. \eeq
  The last
 approximate equality follows from the fact that the fluctuations (28) are so
 small. \\  Thus we
 have for large $\bar{N}$ \beq x \equiv \sigma E_P \beta= 2 t \sqrt{\bar{N}}
 =2t \frac{Mc^2}{\sigma E_P}~. \eeq
  Finally we come to the
  entropy \beq S/k_B = \ln Z_i+\beta U~~, \eeq
  for which we get the exact result
 \beq S/k_B=\frac{x^2}{4t}+ \ln x -\frac{3}{2}\ln t + \ln(\sqrt{\pi}/(2e))~~,
  \eeq or, using the relation (31) and ignoring terms of order $O(1)$,
  \begin{eqnarray} S/k_B &=& t \bar{N} +\frac{1}{2}
  \ln\bar{N}
   \\ &=&\frac{t}{4\pi
  \sigma^2}\frac{A}{4l_P^2}+\frac{1}{2}\ln (
  \frac{A}{l_P^2})~,  \end{eqnarray}
   where $A= 4\pi R_S^2= 16
  \pi G^2M^2/c^4$ is the area of the hori\-zon with Schwarzschild radius $R_S$
  and $l_P^2=G\hbar/c^3$ the Planck length squared. \\ For
  $\sigma^2=t/(4\pi)$ the leading term of $S/k_B$ has the
   Bekenstein-Hawking value
  $A/(4l_P^2)$, for $\sigma = 1/2$, as discussed in ref.\ [19], it
   is slightly smaller if $t=\ln 2$
  etc.. \\ Notice that  the  factor $\exp(-x^2/4t)$ - typical for the heat
  equation - is the decisive one for providing the essential features of
   the thermodynamical properties
  just discussed. \\ Up to now we have been dealing mainly with the large
  $\beta$- (low temperature)- behaviour of the system. There are in addition
  some features for small $\beta$ (high tem\-pe\-rature) worth mentioning.
   Whether
  they have any physical significance - e.g.\ for the big bang era - remains
 to be seen. \\ The internal energy $U$ in eq.\ (24) vanishes for $x=x_1
  =\sqrt{2t}$
  or $k_BT_1 = \sigma E_P/\sqrt{2t}$. $\overline{E^2}$ becomes negative
  (eq.\ (27)): $\overline{E^2}_1=-\sigma^2E_P^2/t$! The entropy (33) at this
  temperature takes the value $S_1/k_B=(1/2)\ln[\pi/(2et^2)]$ which for
  $t=\ln2$ is equal to $0.092\cdots$! \\ According to eq.\ (27) the quantity
  $\overline{E^2}$ vanishes for $x=x_2 = \sqrt{6t},
  k_BT_2=\sigma E_P/\sqrt{6t}$ $<k_B T_1$, with $U= U_2= 2 \sigma
   E_P/\sqrt{6t},
  S_2/k_B =(1/2)\ln[3e\pi/(2t^2)]>S_1/k_B$. \\ Then there is the
   value $x=x_0(t)$ for
  which the entropy (33) vanishes. The resulting equation \[x_0^2+4t
  \ln(\frac{\sqrt{\pi}\,x_0}{2\,e\,t^{3/2}})=0\] cannot
   be solved for $x_0(t)$ explicitly, but the inequality $x_0(t)<
   2\,e\,t^{3/2}/\sqrt{\pi}$ follows immediately. \\ Strange things
    seem to happen
   at temperatures of the order of the Planck energy!

  \section{Remarks}
  The most surprising  feature of the "canonical" quantum
  statistical
  mechanics of the level spectrum (2) is that the expected thermodynamical
  properties of a Schwarzschild black hole are associated with the {\it
  imaginary} part of the partition function for real $g>1$! \\ Superficially
  this appears to question the canonical approach   to the
  thermodynamics of black holes as possibly inappropriate.
   On the other hand the
  thermodynamical properties associated with $Z_i$ are too intriguing and too
  interesting in order to dismiss them. The formal reasons for preferring
  $Z_i$ compared to $Z_r$ as the partition function relevant for the
  thermodynamics of the system have already been stressed in connection with
  the eqs. (22) and (23). The "physical" thermodynamical consequences fully
  justify the more formal conclusions! \\ There is in addition an interesting
  heuristic consistency argument why the imaginary part (23) of
   the partition function is
  the physically relevant one: Suppose the sum (3) does not extend
  up  to $\infty$
  but up to a very large number $N$. Then we get
  \beq Z_N(t,x)= \frac{x}{2\sqrt{\pi}}\int_0^{\infty}\frac{d\tau}{\tau^{3/2}}
  e^{\ds -x^2/(4\tau)}\frac{1-e^{\ds N(t-\tau)}}{1-e^{\ds (t-\tau)}}~ \eeq
  instead of eq.\ (14)
  and the integrand is not singular anymore for $\tau =t$ (the fraction
  resulting from the finite geometrical series just has the value $N$ for
  $\tau =t$) and there is no cut!
 If we now exploit eq. (31) and replace the $x^2$ under the integral sign
  by $4t^2 N$ we get  \begin{eqnarray} Z_N & =&\frac{x}{2\sqrt{\pi}}
  \int_0^{\infty}e^{\ds N
  h(\tau)}k(\tau)~, \\ & &h(\tau)=t-\tau-\frac{t^2}{\tau}~,~~~
 k(\tau)= \frac{1}{\tau^{3/2}}\frac{e^{\ds- N(t-\tau)}-1}{1
 -e^{\ds (t-\tau)}}~.
 \nonumber \end{eqnarray} Evaluating the in\-te\-gral for large $N$ by means of a
 saddle point approximation\cite{43} yields a sad\-dle point for
  $\tau=t$ and the
 result \beq Z_N \approx
 \frac{x}{2t}\sqrt{N}e^{\ds -tN}=\frac{x^2}{4t^2}e^{\ds - x^2/(4t)}~~, \eeq
 where again the relation $x^2=4t^2 N$  has been used. \\ The approximation
  (38) is
 not equal to the function (23) and not a solution of the heat
  equation - there is mainly
 one additional factor $x/\sqrt{t}$ due to the approximations
  involved - but it
 contains the most essential factor $\exp(-x^2/(4t))$ which is so
  important for
 the qualitative structure of the thermodynamics.
  \\Furthermore, the smallness of the
  thermal fluctuations (28) show that  the thermal interactions of the
  black hole with the heat bath are small and, therefore, a "canonical"
  statistical treatment seems plausible and may not be too far off a
  microcanonical one. \\ In any case, the above "canonical" results have to
  be interpreted as requirements on the properties of a heat bath if it is to
  be in thermal equilibrium with the black hole. \\If the whole approach
  discussed above is not unsound then the level spectrum (2) has to be taken
  seriously  and
 a convincing justification of its validity
  is desirable\cite{44}. \\ A final remark: The thermodynamics discussed
   would  be quite different if we
  would not interpret $ g= \exp(t)$ as a fixed number but as the temperature
  dependent fugacity $g=z=\exp(\mu \beta)$, $\mu$: chemical potential, of a
  grand canonical ensemble\cite{45}.
  \\ \\ I thank C.\ Gutsfeld, G. Roepstorff, F. Schramm, T. Strobl and H.\
  Wissowski for help\-ful discussions and critical remarks. Especially the
  numerical tests of C.\ Gutsfeld and F.\ Schramm together with the joint
  discussions including T.\ Strobl prevented me to expect too much
   from the real
  part $Z_r(t,x)$, eq.\ (22), and stimulated me to take the imaginary
   part $Z_i$
  more
  seriously. \\ Finally I thank my wife Dorothea for her support,
  her understanding and her
  patience while this paper was being prepared.
  \\ \\ {\em Note added}: \\ After this paper was submitted as an e-print
  M.\ Perry kindly drew my attention to refs.\cite{46,47} where the imaginary
  part of the partition function is related to the metastable states of
  the system. Especially the last paper[47] is of considerable interest in the
  context of the present article.
    
      \end{document}